# Energy density balance during shock wave implosion in water

Sergey G. Chefranov [*)], Yakov E. Krasik, and Alexander Rososhek

Physics Department, Technion, Haifa 32000, Israel

[*)] csergei@technion.ac.il

**Absract**

Analytical modeling of the evolution of cylindrical and spherical shock waves (shocks) during an implosion in water is presented for an intermediate range of convergence radii. Up to now this range is determined only in experiments observations as a range of implosion radii which are already far from the piston influence, but yet not described by well-known self-similar solutions. The model is based on an analysis of the change in pressure and kinetic energy density, as well as on the corresponding fluxes of internal and kinetic energy densities behind the shock front. It shows that the spatial evolution of the shock velocity strongly depends on the initial compression, the adiabatic index of water, and the geometry of convergence. The model also explains the transition to a rapid like self-similar increase in the shock velocity at only a certain radius of the shock that is observed in experiments. The dependence of the threshold radius, where the shock implosion follows the power law (quasi self-similarity), on the initial compression is determined. It is stated that in the entire range of the shock radii the internal and kinetic energy density fluxes are equal, which is in agreement with known experimental data.

## I. Introduction

The subject of the generation and implosion of converging strong shock waves (shocks) attracts continuous attention because of its importance in high energy density physics and the field of warm dense matter.[1,2] For extremely high pressure shocks, a self-similar approach[3,4] is frequently used to describe the temporal and spatial evolution of spherical and cylindrical shocks. This approach considers convergence without accounting for the piston dynamics used to generate the shock. Its results show a continuing increase in the shock velocity and density behind the shock front along the path of its implosion. In publications [3,4] it is noted that the radial distribution of density, pressure, and flow velocity behind the shock front depends strongly on the adiabatic index and also that the self-similar approach can be applied only for a limited range of converging radii that are far away from piston.

In laboratory experimental conditions, it is seen that the piston evolution can be an important factor that not only determines the parameters of the generated shock but can also influence shock convergence. Recent research [5,6] showed that underwater electrical explosion of cylindrical wire arrays results in the generation of highly uniform converging shocks leading to pressures $\geq 10^{10}$ Pa, densities $\geq 2$ g/cm$^3$, and temperatures of a few eV in the vicinity of the implosion axis.

Experiments and one-dimensional numerical hydrodynamic simulations coupled with the equation of state (EOS) for water, taking into account the temporal dynamics of the radial expansion of cylindrical[7] or planar wire arrays[8], additionally showed that the propagation of the shock is almost independent of the expansion of the array as far as a distance of ~2 mm from it. Thus, one can assume that starting from that radius the shock propagation should be in agreement with a self-similar model. However, the temporal behavior of the shock velocity obtained by shadow streak imaging, as well as by one dimensional numerical simulation, showed a large discrepancy with the self-similar solution. Namely, the velocity of a converging cylindrical shock decreases fast from its maximum value at the vicinity of the expanding array to values slightly exceeding the velocity of sound (see Fig.3). The velocity increases gradually only from a radius of $R \sim 1$ mm relative to the axis, and at $R < 0.1$ mm it becomes close to that predicted by self-similarity.

In this paper, the shock convergence is analyzed using an approach based on comparisons between the energy transferred by the shock to the kinetic and internal energies of the incoming water flow and the potential energy (pressure) behind the shock front. The relations describing the shock velocity evolution during spherical or cylindrical shock implosion observed in the experiments are obtained.

In the next section 2, based on the consideration of the laws of conservation of mass and momentum at the shock wave front in the form of Rankin-Hugonio relations and the isentropic equation of water state, integral relations characterizing the change in pressure and kinetic energy density of the medium behind the front of converging spherical and cylindrical shock waves in water are obtained. At the same time, the analytical dependence on the parameters of the problem for the threshold value of the implosion radius is determined, starting from which the increment of the kinetic energy density of the motion of the medium behind the shock wave front begins to exceed the increment of pressure. In section 3, a similar study is conducted on the basis of consideration of changes in the fluxes of kinetic, potential and internal energy density during implosion. Section 4 presents the main conclusions obtained in this work.

## II. Evolution of kinetic energy density and pressure of the water flow behind the converging shock front.

In this section, we compare the kinetic energy density of the water flow and the pressure behind the front of converging spherical and cylindrical shocks. At the same time, we will make this comparison for two different compression radii corresponding to two consecutive moments of compression time sufficiently distant from the moment of the underwater explosion generating a converging shock wave in the water according to the definition given in the Annotation for the intermediate range of compression radii (see also Fig. 3 and comments before it). We will characterize by the subscript 1 the time of implosion and the corresponding compression radius, which is relatively closer to the time of the underwater explosion compared to the time and the corresponding compression radius, denoted by the subscript 2. Let us consider a cylindrical or spherical shock converging in water so that its front is at radii $R_1$ and $R_2 < R_1$ at times $t_1$ and $t_2 > t_1$, respectively. The density of the water behind the shock front at $t_1$ and $t_2$ is $\rho_1$ and $\rho_2$, respectively, and the normal density of water is $\rho_0$. Let us introduce the dimensionless parameters $\delta_1 = \rho_1/\rho_0$; $\delta_2 = \rho_2/\rho_0$; $x = R_1/R_2$ and $y = \delta_2/\delta_1 = \rho_2/\rho_1$.

Using the mass conservation law (Rankine–Hugoniot conditions) at the shock front[3] $U_a = D_a(\delta_a - 1)/\delta_a$, where $D$ and $U$ are the velocities of the shock front and the water flow behind the front, respectively, and $a = 1, 2$ are the indexes related to times $t_1$ or $t_2$, one obtains

$$\frac{U_2}{U_1} = q\frac{(y\delta_1 - 1)}{y(\delta_1 - 1)}, \tag{1}$$

where $q = D_2/D_1$. The energy density that the shock transfers to the kinetic energy, $\Delta E_K$, of the compressed water flow behind its front during $\Delta t = t_2 - t_1$ can be written as

$$\Delta E_K = \rho_2 U_2^2/2 - \rho_1 U_1^2/2 = \frac{\rho_1 U_1^2}{2}\left(\frac{q^2(y\delta_1 - 1)^2}{y(\delta_1 - 1)^2} - 1\right). \tag{2}$$

To determine the function $y(q)$ on density, a polytropic EOS for water $p_a - p_0 = (K_0/n)(\delta_a^n - 1)$,[3,9] where $K_0 = 2.2 \times 10^9 \, Pa$; $p_0 \approx 10^5 \, Pa$ and $n = 7.15$, is applied. The applicability of this EOS with $n = 7.15$ is limited by pressure $<2\times10^{10}$ Pa; hence, this value of the pressure is used as an upper bound in our estimations. For pressures outside of this range, the value of the exponent $n = 1 + 2(\ln \delta)^{-1} \ln(c/c_0)$ can be determined using the well-known [10] dependence of the local speed of sound $c$ on the compression value $\delta$.

For a strong shock, when $p \gg K_0/n \gg p_0$, one obtains

$$y(q) = (p_2/p_1)^{1/n}. \tag{3}$$

Using the Rankine–Hugoniot condition[3] $p_a - p_0 = \rho_0 D_a^2(\delta_a - 1)/\delta_a$, for $p_a \gg p_0$, the ratio of pressures in the water behind the shock front at $t_1$ and $t_2$ is

$$\frac{p_2}{p_1} = q^2 \frac{(y\delta_1 - 1)}{y(\delta_1 - 1)}. \tag{4}$$

Now, applying Eqs. (3) and (4), function $y(q)$ i.e., the ratio of water compressions as a function of the ratio of shock velocities at $t_1$ and $t_2$, can be written as

$$y^{n+1}(\delta_1 - 1) = q^2(y\delta_1 - 1). \tag{5}$$

It can be shown that, for $n\delta_2 \gg 1$, Eq. (5) reads

$$y = [\delta_1/(\delta_1 - 1)]^{1/n} (D_2/D_1)^{2/n} = d^{1/n} q^{2/n}, \tag{6}$$

where $d = \delta_1/(\delta_1 - 1)$ when $\delta_2 > \delta_1$ and $d = 1$ if $\delta_2 \leq \delta_1$. Finally, using Eqs. (2) and (6), the change in the kinetic energy density (2) of the water flow behind the shock front at radii $R_1$ to $R_2$ is

$$\Delta E_K = \frac{\rho_1 U_1^2}{2} \left[ q^{2\frac{(n-1)}{n}} \frac{\left(q^{2/n} \delta_1 d^{1/n} - 1\right)^2}{(\delta_1 - 1)^2 d^{1/n}} - 1 \right]. \tag{7}$$

The change in the potential energy density, associated with the change in the pressure during $\Delta t = t_2 - t_1$, can be written using Eq. (6) as

$$\Delta E_P = p_2 - p_1 = p_1(q^2 d - 1), \tag{8}$$

where $p_1 = \rho_0 U_1^2 d$. Using Eqs. (7) and (8), one can write the condition for $\Delta E_P \geq \Delta E_K$:

$$\frac{2}{(\delta_1 - 1)}(q^2 d - 1) - q^{\frac{2(n-1)}{n}} \frac{\left(q^{\frac{2}{n}} \delta_1^{\frac{n+1}{n}} - (\delta_1 - 1)^{\frac{1}{n}}\right)^2}{\delta_1^{\frac{1}{n}}(\delta_1 - 1)^{\frac{2n+1}{n}}} + 1 \geq 0. \tag{9}$$

The numerical solution of Eq. (9) allows one to find the range of ratio values $q = D_2/D_1$ versus the initial compression of water $\delta_1$ when the condition for the excess of the potential energy density above the kinetic energy density is satisfied. Further, using $D_1 \approx \left[ p_1 \delta_1 / \rho_0 (\delta_1 - 1) \right]^{0.5}$, one obtains possible values of $D_2 = qD_1$ and, consequently, $\rho_2$ and $p_2$. In Fig. 1, we present the solution of Eq. (9) for $\delta_1$ within the range 1.1–1.85. One can see that when the potential energy density is larger than the kinetic energy density, the range of $q$ is limited by the values of $q_{max}$ and $q_{min}$ with the corresponding values of $D_2^{min}$ and $D_2^{max}$. However, taking into account the applicability of EOS for water limited by $<2 \cdot 10^{11}$ Pa, the maximal possible range of $q$ and $D_2$ is limited by $D_2^{max} \approx 7400 \; m/s$ and $q_{max} \approx 4.8$. These limitations imply a discontinuity in $D_2^{max}$ and $q_{max}$, when the limits are reached (see Fig. 1). Note, that restriction $q > q_{min}$ that follows from Eq. (9) (see Fig. 1) should not be taken into account, since $q_{min} < 1$. Instead, only the condition $q > 1$ must be satisfied because of the applicability of the relation in Eq. (6) used in deriving the relation in Eq. (9).

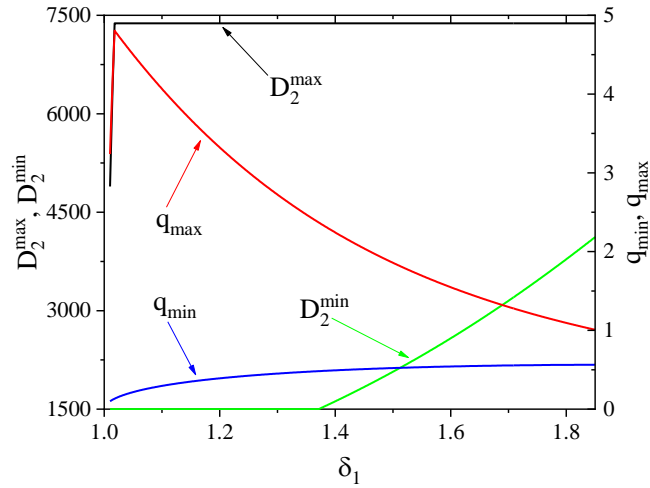

Fig.1. Dependences of $D_1$, $D_2$ and possible range of these velocities ratio limited by $q_{max}$ and $q_{min}$ versus value $\delta_1$.

The curve corresponding to the lower limit of the permissible velocity of the shock wave by inequality (9) begins at 1500 m/s, which corresponds to the speed of sound in water. Since the velocity of the shock wave cannot be less than the speed of sound, the corresponding restriction following from (9) is given only for compression values for which the lower bound of the velocity of the shock wave exceeds the speed of sound in water.

### III. Energy flux balance

In this section, we investigate the threshold shock radius starting from which the shock velocity begins to increase rapidly as the radius of the shock decreases. Let us consider the temporal evolution of the energy density flux behind the front of the converging shock at two different radii. The energy balance equation at these two locations is given by

$$\frac{dE}{dt} = -\Delta J_P - \Delta J_K - \Delta J_I =$$
$$= 2^g \pi R_1^g U_1 \left( p_1 + \rho_1(\varepsilon_1 + \frac{U_1^2}{2}) \right) - 2^g \pi R_2^g U_2 \left( p_2 + \rho_2(\varepsilon_2 + \frac{U_2^2}{2}) \right). \quad (10)$$
$$E = 2^g \pi \int_{R_2}^{R_1} dr\, r^g \left[ \rho \frac{U^2}{2} + \varepsilon \rho \right]$$

Here, $\varepsilon$ is the internal energy density, $g = 1$ for the cylindrical and $g = 2$ for the spherical cases. The expressions in the parentheses on the right hand side of Eq. (10) represent the work produced by the pressure in a unit of time and the fluxes of the internal and kinetic energy in a unit of volume, at the radii $R_1$ and $R_2$. The total energy of the liquid, $E$, in the volume between $R_1$ and $R_2$ is also presented in Eq. (10). Taking into account the Rankine–Hugoniot relations[3] $p_1 = \rho_0 U_1^2 d$, one can write the changes in the fluxes of the potential, internal, and kinetic energy density behind the shock front at radii $R_1$ and $R_2$ as

$$\Delta J_P = \frac{2^g \pi R_1^g \rho_1 U_1^3}{(\delta_1 - 1)} \left( x^{-g} \frac{U_2}{U_1} \frac{p_2}{p_1} - 1 \right) \quad (11)$$

$$\Delta J_K = \frac{2^g \pi R_1^g \rho_1 U_1^3}{2} \left( x^{-g} y \frac{U_2^3}{U_1^3} - 1 \right) \quad (12)$$

$$\Delta J_I = \frac{2^g \pi R_1^g \rho_1 U_1^3}{2} \left( x^{-g} \frac{U_2}{U_1} \frac{p_2}{p_1} \frac{(y\delta_1 - 1)}{(\delta_1 - 1)} - 1 \right) \quad (13)$$

In Eq. (13), we used an approximation in the Rankine–Hugoniot relation[3], namely, $\varepsilon_a \rho_a = 0.5 p_a (\delta_a - 1)$, which can be considered reasonable for $p_a \gg p_0$ and $\varepsilon_a \gg \varepsilon_0$, where $\varepsilon_0$ is the internal energy at normal conditions. Let us note that, using Eqs. (1) and (4), one obtains the equality $\Delta J_K = \Delta J_I$. The latter agrees with the experimental data ($P \leq 4.45\times10^{10}$Pa, $T \leq 3420$ $^0$K, $\rho \leq 2.287$ g/cm$^3$) presented in [10]. Let us determine the condition for which the value of the energy losses associated with the flux of the potential energy density is smaller than the losses associated with the sum of the internal and kinetic energy density fluxes $\Delta J_P < 2\Delta J_K$. Using Eqs. (6), (11), and (12) this condition reads

$$u^{\frac{3n-4}{2}}(u-1)^2(2-u) < x^g(2-\delta_1)\delta_1^{\frac{3n-1}{2}}(\delta_1-1)^{1/2}$$
$$u \equiv \delta_2 = \delta_1\left(\frac{\delta_1}{\delta_1-1}\right)^{1/n}\left(\frac{D_2}{D_1}\right)^{2/n} = \delta_1 d^{1/n} q^{2/n}$$
, (14)

or in explicit form

$$\delta_2^{\frac{3n-4}{2}}(\delta_2-1)^2(2-\delta_2) < \left(\frac{R_1}{R_2}\right)^g (2-\delta_1)\delta_1^{\frac{3n-1}{2}}(\delta_1-1)^{1/2}.$$

Let us consider condition in Eq. (14) for large water compression, which can be realized in the vicinity of the shock implosion axis or origin, when $\delta_2 > \delta_1 > 2$ in the limit $\delta_2 \gg 2$.

$$\delta_2^{\frac{3n+2}{2}} > x^g \delta_1^{\frac{3n-1}{2}}(\delta_1-2)\sqrt{\delta_1-1}.$$

Let us recall here that the polytropic EOS with adiabatic index $n = 7.15$, which is valid in the range of pressure behind the shock front $P < 2\times10^9$ Pa, cannot be used in this approximation and another index should be applied [9]. This inequality allows one to obtain the possible values of $\delta_2 = f(R_1/R_2)$, as well as of $D_2 = c_0 f_D(R_1/R_2)$, for given values $\delta_1$ at $R_1$ when the condition $\Delta J_P < 2\Delta J_K$ is satisfied. The latter dependence reads

$$D_2 > c_0\left(\frac{(\delta_1^n-1)\delta_1}{n(\delta_1-1)}\right)^{1/2}\left[\frac{(\delta_1-1)^{2n+1}(\delta_1-2)^n}{\delta_1^{3n+1}}\right]^{\frac{1}{3n+2}}\left(\frac{R_1}{R_2}\right)^{\frac{gn}{3n+2}}. \quad (15)$$

Now let us consider the inequality in Eq. (14) when the water compression is moderate, i.e., $1 < \delta_1 < \delta_2 < 1.85$. As an example, the dependencies of the boundary values of $D_2$ and $\rho_2$ versus the radii ratio $R_1/R_2$ for $\rho_1 = 1.13$ are shown in Fig. 2 for cylindrical converging shocks together with the results of one dimensional hydrodynamic simulations. One can see that the results of simulations satisfy inequality (14) with the entire considered range of $R_1/R_2$.

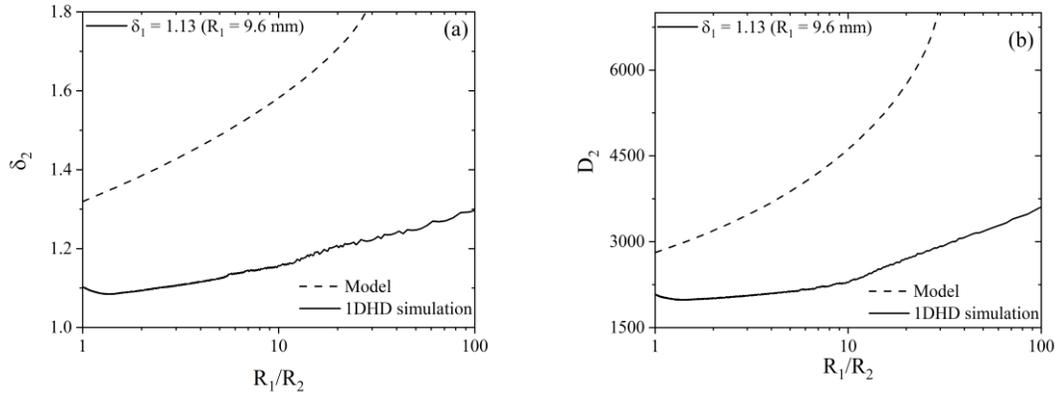

Fig. 2. Dependences of $\delta_2(x = R_1/R_2)$ and $D_2(R_1/R_2)$ for cylindrical shock convergence.

Now let us obtain the sufficient condition for the fulfillment of inequality in Eq. (14). The left hand side of Eq. (14), $f(u) = u^{\frac{3n-4}{2}}(u-1)^2(2-u)$, has a maximum $f_{max} \approx 23.26$ for $u_{max} \equiv \delta_2 \approx 1.858$. Thus, the condition $\Delta J_P < 2\Delta J_K$ will always be satisfied for the range of radii ratio:

$$x > x_{min} = \left( \frac{23.26}{\delta_1^{\frac{3n-1}{2}}(2-\delta_1)\sqrt{\delta_1-1}} \right)^{1/g} \geq 1 \quad (16)$$

For example, for $\delta_1 = 1.3$, for a cylindrical shock one obtains $x_{min} \approx 4.15$ and for a spherical shock $x_{min} \approx 2.04$. The obtained estimate for a cylindrical shock wave is consistent with the simulation results shown in Fig. 3, simulating the non-monotonic dependence of the velocity of a converging cylindrical shock wave in water on the radius of the shock wave observed in the experiment. Indeed, according to Fig.3, when the compression value $\delta_1 \approx 1.31$ is close to the value $\delta_1 = 1.3$, starting from the compression radius $R_2 \approx 0.8$ mm, there is a sharp increase in the velocity of the converging cylindrical shock wave. On the other hand, it follows from Fig.3 that the beginning of the decrease in the velocity of the shock wave during compression occurs at the magnitude $R_1 \approx 3.3$ mm. Starting from this compression radius, the noticeable recharge of the shock wave energy from the piston stops, which corresponds to the upper limit of the intermediate scale range considered in this paper. At the same time, the value of the ratio $R_1/R_2 \approx 3.3/0.8 \approx 4.13$ is close to the estimate $x_{min} \approx 4.15$ obtained from Eq. (16).

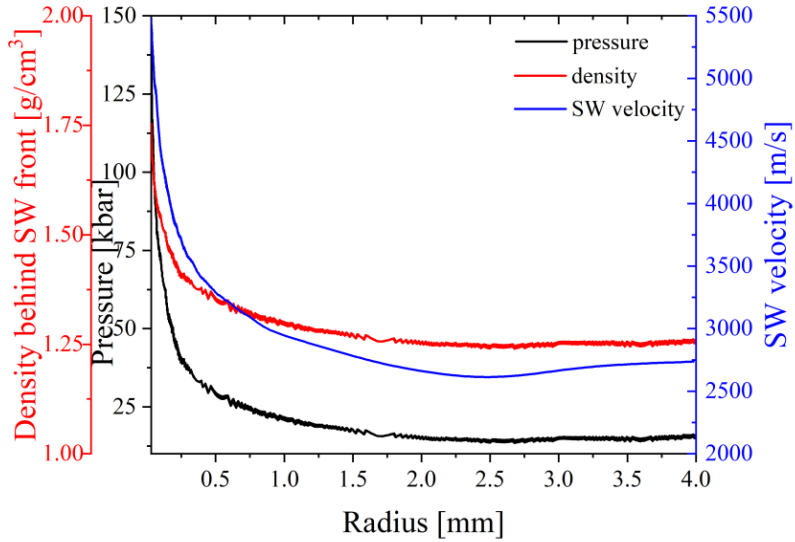

Fig. 3. Radial dependences of the pressure and density behind a shock and the shock velocity.

Finally, let us consider the condition for a decrease in the power *dE/dt* [see Eq. (10)] when the potential energy flux has negative value $\Delta J_P = -|\Delta J_P| < 0$, which can be realized when, because of the dominating geometrical factor, the potential energy density flux will be increased during the shock implosion. According to equality in Eq. (11) and taking into account the Rankine–Hugoniot relations, the condition $\Delta J_P < 0$ reads $x^g > \dfrac{U_2 p_2}{U_1 p_1} = \left(\dfrac{p_2}{p_1}\right)^{3/2} \left(\dfrac{y\delta_1 - 1}{y(\delta_1 - 1)}\right)^{1/2}$. Now, using equality in

Eq. (3) obtained from the polytropic EOS for water, this inequality can be re-written as

$$x^g > y^{\frac{3n-1}{2}} \left(\frac{y\delta_1 - 1}{\delta_1 - 1}\right)^{1/2} \; ; y = \delta_2/\delta_1 \; . \tag{17}$$

One can see that the dependence of the threshold value of $\delta_{2th}(x, \delta_1)$ on $x = R_1/R_2$ for different values of initial compression $\delta_1$ is similar to that in Fig. 2. This indicates that inequality in Eq. (17) is consistent with the simulation data in the corresponding range of changes in the radii of a converging cylindrical shock.

Using Eqs. (1), (4), and (6), conditions $\Delta J_P < 0$ and $\Delta J_I > 0$ in Eqs. (11) and (13) result in inequality for the radii ratio (when in (17) $\delta_2 > \delta_1$ and in addition the limit $\delta_2 \gg 1$ is considered):

$$x_{2th} \approx \left(q^{\frac{3n+2}{n}} d^{\frac{3n+1}{n}}\right)^{1/g} > x > x_{1th} \approx \left(q^3 d^2\right)^{1/g} . \tag{18}$$

Here, the lower limit $x > x_{1th}$ follows from the condition $\Delta J_P < 0$ stated in Eq. (17) and the upper limit $x < x_{2th}$ follows from the condition $\Delta J_I > 0$. Note that, when defining $x > x_{1th}$, one can use only the values $q = D_2/D_1 > 1$, which corresponds to values $y = \delta_2/\delta_1 > 1$ because for the case $y < 1$ in Eq. (18) $d = 1$ must be taken equality (see derivation of Eq. (6)) ,

When taking into account Eqs. (1), (4), and (6) and assuming that $n\delta_2 \gg 1$, one can write

$$x_{1th} = \left(\frac{D_2}{D_1}\right)^{3/g} \left(\frac{\delta_1}{\delta_1 - 1}\right)^{2/g} \tag{19}$$

$$x_{2th} = \left(\frac{D_2}{D_1}\right)^{\frac{3n+2}{ng}} \left(\frac{\delta_1}{\delta_1 - 1}\right)^{\frac{3n+1}{ng}} . \tag{20}$$

Here, let us note that the power degree in Eq. (19) agrees with the self-similar solution presented in [12] for converging cylindrical and spherical shocks. Indeed, the threshold ratio $D_2/D_1$ in (19), which corresponds to the threshold value $R_1/R_2$, reads

$$\frac{D_2}{D_1} = \left(\frac{\delta_1 - 1}{\delta_1}\right)^{1/3} \left(\frac{R_1}{R_2}\right)^{g/3} \qquad (21)$$

The self-similarity parameter $\alpha$ [3] is $\alpha = 3/(3+g)$. Thus, for a cylindrical shock $\alpha = 0.75$ and for a spherical shock $\alpha = 0.6$. These values of $\alpha$ are the same as those presented in [12]. The same consideration used for Eq. (20) gives

$$\frac{D_2}{D_1} = \left(\frac{\delta_1 - 1}{\delta_1}\right)^{\frac{3n+1}{3n+2}} \left(\frac{R_1}{R_2}\right)^{\frac{gn}{3n+2}}. \qquad (22)$$

From Eq. (22), one obtains $\alpha = \dfrac{3n+2}{2+n(g+3)}$, which gives $\alpha = 0.766$ and $\alpha = 0.621$ for cylindrical and spherical shocks, respectively. These values of parameter $\alpha$ are also close to those presented in [12]. Note that the power laws in Eq. (21) and Eq. (22) are dependent on the initial compression value $\delta_1$. The latter differs from the self-similar solution[3,4] describing a converging shock far away from the piston, when the information about the initial conditions is already lost. For the shock convergence described by Eq. (21) and Eq. (22) (let us define it as a quasi-self-similar power law), this information is preserved, and therefore, this law describes an intermediate stage of shock implosion between the known self-similar regime and a solution that completely depends on the specific dynamics of the piston. However, in Eq. (21) and Eq. (22), not the compression modes themselves are presented, but only the upper and lower bounds of the permissible type of dependence of the velocity of a converging spherical or cylindrical shock wave on its radius, when the necessary condition for an energetically optimal stable regime of purely radial motion of the medium behind the shock wave front is met [13], [14]. Thus from Eq. (18) it is possible to obtain inequality:

$$D_1\left(\frac{\delta_1-1}{\delta_1}\right)^{\frac{3n+1}{3n+2}}\left(\frac{R_1}{R_2}\right)^{\frac{gn}{3n+2}} < D_2 < D_1\left(\frac{\delta_1-1}{\delta_1}\right)^{\frac{2}{3}}\left(\frac{R_1}{R_2}\right)^{\frac{g}{3}} \qquad (23)$$

The shock threshold radius $R_{2th}$, starting from which $D_2 > D_1$, satisfies the range of values limited by Eq. (23) and, respectively a negative potential energy flow, $\Delta J_P < 0$, can be determined using the condition that the right hand side of Eq. (23) should be larger than $D_1$:

$$R_2 < R_{2th} \equiv R_1\left(\frac{\delta_1-1}{\delta_1}\right)^{2/g}. \qquad (24)$$

Let us consider inequality in Eq. (24) for $R_1 \approx 4$ mm and $\delta_1 \approx 1.256$, which correspond to the initial conditions of simulations, the results of which are presented in Fig. 3. In this case, for a cylindrical shock one obtains $R_2 < R_{2th} \approx 0.166$ mm. When this condition is fulfilled, the right hand side of inequality in Eq. (18) is satisfied; i.e., one obtains negative pressure flow $\Delta J_P < 0$ in the energy balance Eq. (10). The shock threshold radius obtained from (24) in fact corresponds to the end of the transition region (see Fig. 3) for the velocity of a converging cylindrical shock.

## IV Conclusion

We presented an analytical model of the evolution of converging cylindrical and spherical shocks in water, based on the comparison of the potential, internal, and kinetic energy densities of the flow behind the shock front. It was found that the internal and kinetic energy density fluxes, which determine the decrease in the total energy of the flow behind the shock front, are equal, in agreement with the experimental data [10], obtained in earlier research for pressures $p \leq 4.5 \times 10^{10}$ Pa. It was shown that the evolution of the implosion velocity depends strongly on the initial compression of the water flow and the geometry of the convergence. Namely, the shock does not necessarily increase its velocity as the radius decreases due to the shock instability [13] and [14]. Here we stated that there is a range of initial water compressions values that results in the decrease in the shock velocity along some path of its convergence for any relevant value of the adiabatic index in the polytropic EOS of water. Finally, initial threshold radii corresponding to water initial compression values were found that allow quasi-self-similar shock propagation with quasi-self-similarity parameter values in agreement with earlier research.

## Acknowledgments

We thank Dr. A. Velikovich, Dr. V. Gurovich for fruitful discussions and reading of this manuscript. This research was supported by the Israeli Science Foundation Grant No. 492/18.

## Data availability

The data that support the findings of this study are available from the corresponding author upon reasonable request.